\documentclass[twocolumn,showpacs]{revtex4-1}
\usepackage{graphics,epsfig,graphicx}
\usepackage{color}

\usepackage{latexsym}
\usepackage{calrsfs}
\usepackage{changes}

\begin{document}



\title{On-demand confinement of semiconductor excitons by all-optical control}

\author{Mathieu Alloing$^{1}$, Aristide Lema\^{i}tre$^{2}$, Elisabeth
Galopin$^{2}$ and Fran\c{c}ois
Dubin$^{1}$}

\affiliation{$^{1}$ ICFO-The Institut of Photonic Sciences,
Av. Carl Friedrich Gauss, num. 3, 08860
Castelldefels (Barcelona), Spain}
\affiliation{
$^{2}$ Laboratoire de Photonique et Nanostructures, LPN/CNRS, Route de
Nozay, 91460 Marcoussis, France}

\date{\today }

\begin{abstract}

In condensed-matter physics, remarkable advances have been made with atomic systems by
establishing a thorough control over cooling and trapping techniques\cite{Bloch_2008}.
In semiconductors, this method may also provide a
deterministic approach to reach the long standing goal of harnessing collective quantum phenomena with exciton gases.  
While long-lived excitons are simply cooled to very low temperatures using 
cryogenic apparatus, engineering confining potentials has been a
challenging task. This degree of control was only achieved recently with devices realized by highly demanding
nano-fabrication processes\cite{Timofeev_2008,High_2008,High_2009, Kotthaus_2012}. Here, we demonstrate an alternative to this
technology and show how a proper optical excitation allows to manipulate in-situ
the exciton transport. Our approach is based on the
optically controlled injection and spatial patterning of charges trapped in a
field-effect device. Thus, electric field gradients are created and implement microscopic traps or anti-traps for the excitons dipole. Accordingly, any
confinement geometry can be realized by shaping the spatial profile of a laser excitation. Hence,
we succeed in trapping exciton gases in a density range where quantum
correlations are predicted at our very low bath temperature \cite{Ivanov_2004}.

\end{abstract}

\maketitle


In the quest for a model system to engineer collective quantum states in the
solid-state, spatially indirect excitons of bilayer heterostructures provide a remarkable ground\cite{Eisenstein_2004,Butov_2007,Snoke_review,Croxall_2008,Seamons_2009}.
Indeed, these are long-lived boson-like quasi-particles which exhibit a very
unique property: a large and well oriented electric dipole. Consequently, the
excitonic flux can be manipulated by controlling the dipolar interaction between indirect
excitons and external electric fields: Indirect excitons are high-field seekers
and are then efficiently transported or confined by appying a spatially varying electric
field onto a bilayer structure. Recent experiments have shown that this can be achieved with field-effect 
devices where micro-patterned metallic gate electrodes serve to imprint a
well defined electrostatic landscape\cite{Chen_2006,Timofeev_2008,High_2008,High_2009,Kotthaus_2012,
Kotthaus_2007}. Although it is effective, this approach requires to vary the electrode pattern in order to create
different confinement geometries. Here, we show that this constraint can be removed by demonstrating that
the confinement for indirect indirect excitons can be programmed all-optically.
As already shown with various systems\cite{Bloch_2008,Wertz_2010,Amo_2010,Hammack_2006}, the
optical control is a very flexible and easily implementable tool which 
constitutes a significant advantage over the state-of-the-art technology.
Hence, we introduce a complete set of operations to manipulate the
excitonic flux through the realization of trapping or anti-trapping potentials
with microscopic dimensions. These realize a toolbox to engineer at will any
confinement geometry where ultra-cold exciton gases could be manipulated. 

In our experiments, indirect excitons were created in a bilayer heterostructure consisting of
a GaAs/Al$_{0.33}$Ga$_{0.77}$As/GaAs double quantum well (DQW) embedded in a
$n^+-i-n^+$  field-effect device (see Supplementary
Informations). The $n^+$ layers, extending over the
entire structure surface, serve as gate electrodes that are polarized to apply a
uniform
electric field perpendicular to the DQW plane. Thus, minimum energy
states for electrons and holes lie in distinct quantum wells such that the
ground excitonic transition
is spatially indirect (see Figure 1.A).

\begin{figure}
\includegraphics[width=8.5cm]{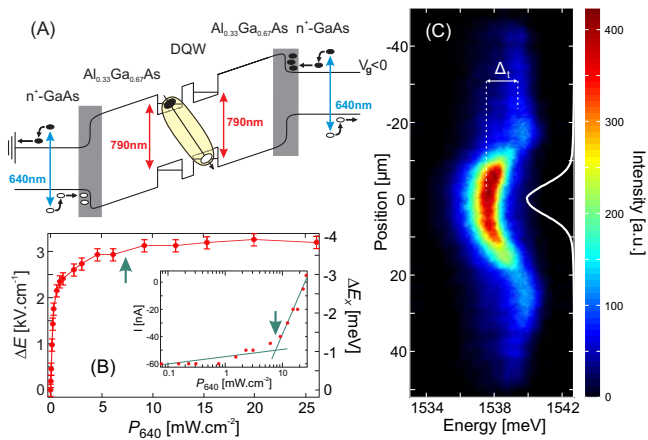}
\caption{(A): Sketch of the energy levels in the field-effect
device embedding a double quantum well (DQW). Photo-injected electrons
and holes are represented by full and open circles respectively. Our top
gate-electrode is biased at V$_\mathrm{g}$= -1.4 V while
the bottom electrode is electrically isolated and grounded. ''Preparation'' and
''writing'' beams are represented by the blue and red arrows respectively.
The grey areas display the heterojunctions where photo-injected carriers
accumulate. (B): Variation of the internal electric field amplitude, $\Delta
E$=($E$-$E^{(0)}$) where $E^{(0)}$= 33 kV.cm$^{-1}$ is the nominal field
amplitude, as a
function of the power of the ``preparation`` beam, $P_{640}$. $\Delta
E_X$=($E_X-E^{(0)}_X$) is the corresponding variation of the energy of
photoluminescence of indirect excitons ($E^{(0)}_X$=1.5418 eV is the emission
energy without ''preparation'' pulse).
The inset shows the variation of the photo-current in these measurements. (C):
Spatially and spectrally resolved emission of indirect excitons loaded by a
probe pulse ( at 8.5 mW.cm$^{-2}$) turned on 500 ns after a ''preparation''
pulse of 10 $\mu$m waist and power equal to 100 mW.cm$^{-2}$ [the white
line displays the profile of the ``preparation`` laser beam].  All measurements
were carried out at 350 mK.}
\end{figure}

Our optical control relies on a scheme of two laser excitations  (Figure
1.A). First, a "preparation" beam creates an out-of-equilibrium population of
charges at the $n^+$-$i$ heterojunctions. This excess of charges modifies the
internal electric field. The purpose of the second "writing" beam is two-fold:
first it depletes \textit{locally} these charges thus producing a lateral
potential gradient yielding traps and anti-traps for indirect excitons. In
addition, it creates dense gases of indirect excitons inside the DQW, as
required to engineer collective quantum phenomena.  The
"writing" beam was produced by a laser diode emitting at $\sim 790$~nm, i.e. at
the energy of the DQW direct excitonic transition. Carriers are then
photo-created directly inside the DQW where indirect excitons are formed after
tunneling towards minimum energy states. The "preparation" beam wavelength was
tuned at 640~nm (from a second laser diode), slightly below the
Al$_{0.33}$Ga$_{0.77}$As band-gap\cite{Bosio_1988}. Hence, only a negligible
part of the light
($\sim 2\%$) is absorbed by the DQW while the largest fraction  is absorbed in
the $n$-doped GaAs electrodes ($\sim 29 \%$ and $\sim 46 \%$ for the top and
back contacts at the right and left hand-side in Fig. 1.A
respectively\cite{Monemar_1976}). The largest part of the carriers
photo-created in the contacts generates a photo-current which
increases with the laser intensity, up to $\sim 50$~nA for the highest
excitation power. The other part produces the out-of-equilibrium population of
charges electrically transported towards the $n^+$-$i$ heterojunctions. There,
potential barriers are formed due to the rectification of potentials between the
$n^+$-GaAs and Al$_{0.33}$Ga$_{0.77}$As layers\cite{Tersoff_1984}. At 350~mK, our bath temperature, the
thermal activation is too small ($k_B T\sim 30$~$\mu$eV) to overcome these
barriers\cite{Alloing_2012}. Thus, charges accumulate and then enhance the electric field inside
the field-effect device\cite{note_h}. 

The resultant internal electric field amplitude  ($E$) was measured as a
function of the power of the "preparation" beam (see Figure 1.B). $E$
was determined from a pump-probe sequence
where a "preparation" pump pulse is followed by a weak excitation probe, from a
third laser beam also tuned at 790~nm (see also the
Supplementary Informations). The probe pulse creates a dilute gas of indirect
excitons in the DQW which acts as a sensitive and non-perturbative probe of
$E$. Indeed, the indirect exciton emission energy
$E_X$ is governed, in the dilute regime, by the interaction (or Stark effect)
between the excitonic dipole ($d\approx e.12$~nm, $e$ being the electron charge)
and the field, such that  $E_X\sim -d.E$.
As shown in Fig. 1.B, the electric field amplitude $E$ is enhanced by the
"preparation" excitation which manifests that charges accumulate on both sides
of
the Al$_{0.33}$Ga$_{0.77}$As barriers. However, $E$ saturates at large pump intensities, 
as the charge accumulation is too important
to be blocked efficiently by the potential barrier at the heterojunctions. This
coincides with a steep increase of the photocurrent (see inset in Fig. 1.B).

We observed a strong lateral localization of excess charges
over long timescales. This phenomenon is evidenced by the spatially and
spectrally resolved emission of indirect excitons, as shown in Fig. 1.C. In
these experiments the "preparation" beam waist was narrowed down to 10~$\mu$m,
while the probe beam waist was much larger, $\sim$ 67~$\mu$m. The probe beam was
turned on 500~ns after the end of the "preparation" pulse, a delay much longer
than
the indirect exciton lifetime ($\sim 138$~ns, see Supplementary Informations).
Then, 20~ns after the end of the probe pulse, the mapping displayed in Fig. 1.C was
acquired.  Under the "preparation" beam excitation area, $E_X$ is shifted to
lower energies by $\Delta_t \sim 1.7$~meV while outside the emission occurs at
the unperturbed energy. This indicates that, first, the charge accumulation
lasts for several hundreds of nanoseconds, and second, that the carriers lateral
diffusion is very slow. Hence, localized excess charges create a confinement potential for indirect excitons. As clearly shown in Fig. 1.C, 
the emission exhibits a much larger intensity under the region excited by the
"preparation" beam. 
Indirect excitons created by the probe pulse explore, on the ns timescale,
the trapping potential created by the ``preparation`` pulse. It has a 
depth $\Delta_t$ almost two orders
of magnitude greater than the thermal activation energy, and as mentioned above,
the
trap is long-lived, with a lifetime greater than the 500 ns time delay between
the ``preparation`` and the probe excitations. Therefore, the slow trapping
dynamics
gives us a unique opportunity to control the excitonic flux, e.g. by optically
imprinting a trapping potential and subsequently injecting excitons inside the trap.

\begin{figure}
\includegraphics[width=8.5cm]{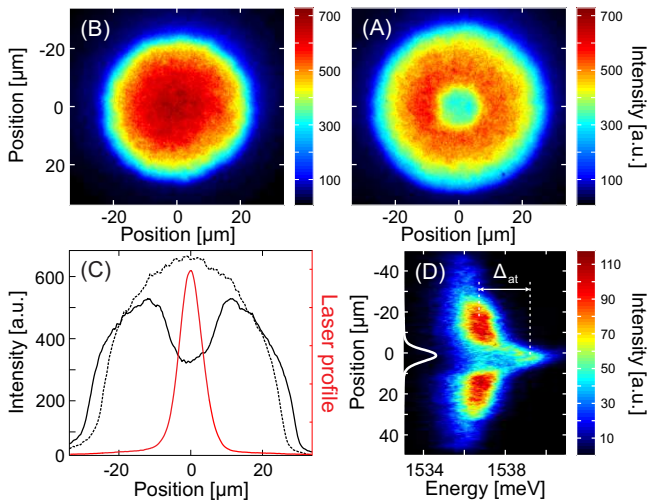}
\caption{(A-B) Photoluminescence emission of indirect excitons for a 8 $\mu$m
extended ``writing`` pulse (at 1.5 W.cm$^{-2}$) with (A) and without (B)
excitation with a
''preparation'' beam of 79 $\mu$m
waist (at 15 mW.cm$^{-2}$). (C): Intensity profiles taken at the center of the
images (A) and (B), solid and dashed lines respectively, together with the
profile of the ``writing`` beam (red line). (D): Spatially and
spectrally resolved emission of a dilute gas of indirect excitons loaded by a
probe pulse (at 4.3 mW.cm$^{-2}$) turned on 500 ns
after the laser sequence used in (A).
$\Delta_{at}\approx$ 2 meV marks the amplitude of the anti-trapping potential
and the white line displays the beam profile of the ''writing'' pulse. All
measurements were carried out at a bath temperature
of 350 mK.}
\end{figure}

While a weak probe excitation does not perturb the field applied onto the
DQW, an intense excitation produced by a ''writing'' pulse can significantly
affect  the spatial profile of the carriers injected by the "preparation" beam.
At the same time the former pulse creates high densities of indirect excitons. 
This point is illustrated in Figure 2 for an experiment where a tightly focused
``writing`` beam (8~$\mu$m waist) was positioned at the center of a wide
"preparation" laser beam (79~$\mu$m
waist). The time delay between the two pulses is again 500~ns. In Figure 2.A,
20~ns after the "writing" beam, the spatially resolved emission of indirect
excitons exhibits a ring-shaped emission pattern. The inner dark region has a
spatial profile reproducing closely the ''writing'' beam shape: In this region
the photoluminescence signal drops by $\sim 50$~\% compared to a reference
situation where no ``preparation`` beam is used (see Fig. 2.C). Incidentally, we
note for this reference case that the emission has a spatial extension
much larger than the writing pulse area itself (see Fig. 2.B):
strong repulsive dipolar interactions between indirect excitons induce a large
exciton diffusion\cite{Remeika_2009,Alloing_2011}.

\begin{figure}
\includegraphics[width=8.5cm]{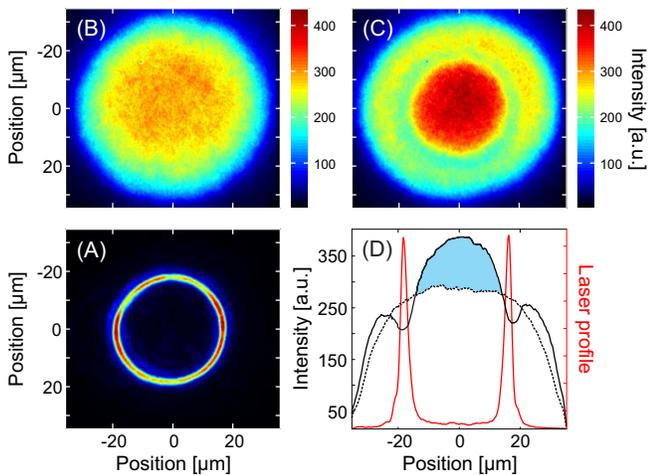}
\caption{(B-C) Photoluminescence emission of indirect excitons for the
ring-shaped ``writing`` beam shown in the panel (A) (18 $\mu$m radius and 3
$\mu$m width at 0.75 W.cm$^{-2}$) without (B) and with (C) excitation of the
field effect device by a ''preparation'' beam 
with ~50~$\mu$m waist (at 3.8mW.cm$^{-2}$).
(D): Intensity profiles taken at the center of the images displayed in (B) and
(C), dashed and solid lines respectively, together with the profile of the
``writing`` beam (red line). The filled area marks the increase of the
photoluminescence when the hollow-trap is imprinted. All
measurements were carried out at a bath temperature of 350 mK.}
\end{figure}

The occurrence of the ring pattern in Fig. 2.A indicates that the central region
is now a low field region, from where the excitons are dragged away. Hence,
trapped carriers have been depleted from this region by the ''writing'' beam.
While
the mechanism responsible for this depletion is not fully established, we shall
underline that the ``writing`` beam has a large intensity (1.5~W.cm$^{-2}$) so
that the laser excitation results in a heating restricted to the illuminated
region \cite{Ivanov_2004,Hammack_2009} (see also the Supplementary
Informations). Thus, the thermal energy can be sufficient to trigger the diffusion of carriers 
in the plane of the heterojunctions.
To confirm and measure the potential step for indirect excitons in the central
region, a third probe beam was used, in the same way as for the experiments
shown in Fig. 1.C. The spatial and spectral mapping of the exciton luminescence
was measured 20~ns after the probe pulse (Fig. 2D). At the ''writing`` beam
position, the exciton energy is larger than in the surrounding, by
$\Delta_{at}\approx$ 2 meV. This shows that the ''writing'' beam has introduced
a potential gradient, dragging the indirect excitons away from the beam
position and thus realizing an anti-trap. Let us note that such a localized
potential barrier appears well suited to control the excitonic flux notably in
transistor architectures\cite{High_2008,Grosso_2009}.

This technique can be easily extended to realize a hollow-trap for indirect
excitons, as we show in the following. Such trap consists of a ring-shaped
potential barrier confining an excitonic gas\cite{Grimm_1998}. To engineer it,
we solely shaped the ''writing'' beam into a ring. 
Again we used a sequenced optical control where a spatially
wide and uniform ``preparation`` pulse is followed by a ring-shaped ''writing``
pulse (see Supplementary Informations). The ring-shaped pattern is shown in
Fig. 3.A. It was obtained by using an axicon lens yielding a ring with a
18~$\mu$m radius and a 3~$\mu$m thickness. First, we present in Fig. 3.B the
spatially resolved emission when the field-effect device is not prepared by the
''preparation'' beam. No trace of the ''writing`` beam shape is observed owing
to the large diffusion of indirect excitons. On the contrary, when the sample is
first prepared by the ``preparation`` beam (Fig. 3.C), the emission, 20 ns after
the end of the ''writing'' pulse, is more intense (same color scale as in Fig 3.B)
and localized within the area defined by the ring shape. The writing beam has, at
the same time, defined a hollow-trap for excitons and built up a population of
indirect excitons which relax rapidly toward the trap center. As shown in Fig.
3.D, this excitation scheme is efficient to create and confine a dense gas of
indirect excitons as the photoluminescence increases by as much
as 30~\% at the center of the trap. From the spectrally resolved
photoluminescence emission (see Supplementary Informations), we estimate that
the exciton concentration is $\sim$ 5 10$^{10}$ cm$^{-2}$ inside the trap. This density lies in the range where 
quantum statistical signatures are predicted at our very low bath
temperature\cite{Ivanov_2004}.

\newpage

\begin{center}
 \LARGE{\textbf{Supplementary Informations}}
\end{center}

\normalfont


\section{Sample details}

We studied a field-effect device which is schematically represented in Figure
S1. It consists of a $n^+-i-n^+$ heterostructure
embedding a double quantum well (DQW) made of two 8 nm wide GaAs quantum wells
separated by a 4 nm Al$_{0.33}$Ga$_{0.67}$As barrier. The DQW is placed between
two
Al$_{0.33}$Ga$_{0.67}$As spacers of 200nm thickness. In our experiments, a
static potential (V$_{g}$) was applied between two n$^{+}$-GaAs layers (Si doped
with $n_{Si} \sim 5.10^{17}$ cm$^{−3}$) which serve as gate electrodes. The top
and back gates are 105nm and 305nm thick respectively. The back gate is
electrically connected with a metallic contact deposited after etching
of the sample. In our experiments, this electrode
is grounded and electrically isolated from the sample holder by a 800nm undoped
GaAs layer incorporating a AlAs/GaAs superlattice. In addition, the top
contact consists of a  mesa with a 500x500 $\mu$m$^2$
aperture through which the
photoluminescence is collected.

\section{Experimental Setup}

The field-effect device was placed on the Helium 3 insert of a closed cycle
Helium 4 cryostat
(Heliox-ACV from Oxford Instruments). An aspheric lens with a 0.6 numerical
aperture is
embedded inside the cryostat in front of the sample and positioned by
piezo-electric transducers
(ML17 from MechOnics-Ag). We optimized the optical resolution of our microscope
by
introducing a mechanical coupling between the Helium 3 insert and the part
holding the
aspheric lens. Thus, the amplitude of mechanical vibrations does not exceed 2
microns (in a
frequency range up to $\sim$ 1 kHz) while the sample can be cooled to
temperatures as low
as 330 mK. In our experiments, the aspheric lens was used to excite the
semiconductor sample but also to collect the photoluminescence that was
reemitted. The latter was then directed to an imaging spectrograph coupled to an
intensified CCD camera (Picostar-UF from La Vision). Thus, we studied the
emission of indirect excitons with a 2 ns time resolution, either in real space
or including
a spectral resolution of 200 $\mu$eV.

\vspace{0.5cm}
\centerline{\includegraphics[width=8.5cm]{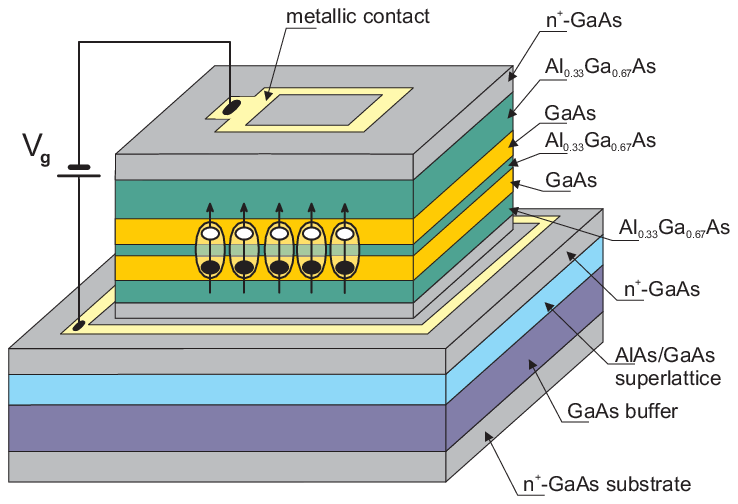}}
\noindent {\bf Fig. S1.} Schematic representation of the semiconductor
sample that we studied at very low temperature (350 mK).

\section{Electric dipole of indirect excitons}

In our experiments, a negative bias ($V_g$= -1.4 V) was applied to the top gate
electrode such that electrons and holes are confined in the bottom and top
quantum wells respectively (see Fig. S1). Thus, the ground excitonic transition
is of indirect type and indirect excitons exhibit an
electric dipole $\textbf{d}=e.\textbf{r}$ oriented perpendicularly to the plane
of the DQW, $e$ being the electron charge and $\textbf{r}$ the distance between
the electron and hole constituting the exciton. In this geometry, indirect
excitons interact strongly with the externally applied electric field
$\textbf{E}$ through a
dipolar interaction $E_{dip}\sim -\textbf{d}.\textbf{E}$. The amplitude of the
electric field in the plane of the
DQW is defined by the applied bias $V_{g}$ and by the thickness of the
heterostructure, $l$. It reads $E=|\textbf{E}|=V_{g}/l$. Therefore, the exciton
electric dipole moment is deduced by varying the applied voltage, $V_g$, while
monitoring
the energy of the photoluminescence emission, $E_X\sim E_{dip}$. In Figure S2 we
present the result of such measurement where we obtained a
variation of $25.0\pm0.4$ meV.V$^{-1}$ which combined with the thickness of our
field-effect device ($l=420$ nm) gives an electric dipole
$d=|\textbf{d}|=e.(10.3\pm0.2)$ nm for indirect excitons. This value is in
good
agreement with the characteristic dimension of the DQW where the center of the
two quantum wells are separated by 12 nm. The latter distance provides the
expected electric dipole moment.

\vspace{0.5cm}
\centerline{\includegraphics[width=8.5cm]{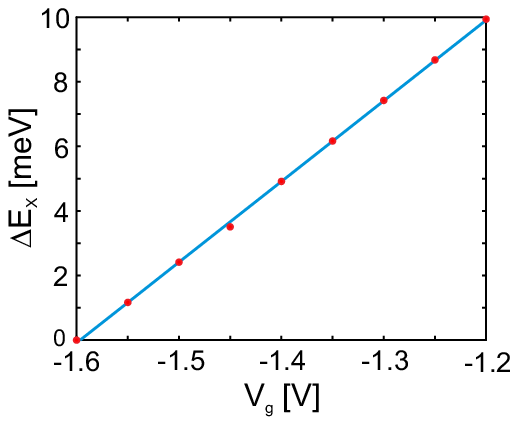}}
\noindent {\bf Fig. S2.} Variation of the energy of the photoluminescence of
indirect
excitons, $\Delta E_X$, as a function of the gate voltage, $V_g$. Measurements
were performed at 350 mK and under a weak (5 mW.cm$^{-2}$) excitation of the DQW
at 790
nm to prevent the photo-injection of free carriers. Measurements were realized
at 350 mK.

\section{Laser induced heating and optical lifetime of indirect excitons}

In Figure S3.A-B we show the spectrally and spatially resolved emission of
indirect excitons at the end (A) and 4 ns after (B) a 300 ns long "writing"
pulse at 10 W.cm$^{-2}$. We note during the laser pulse that the  emission is
weaker
in the region which is illuminated compared to outside. Furthermore, once the
laser excitation is terminated the signal at
the position of the "writing" beam becomes the most intense (see Fig. S3.C).
This behavior is
characteristic of indirect excitons and marks the formation and collapse of the
so-called inner-ring. It signals
a laser induced heating in the region that is illuminated
\cite{Ivanov_2004,Hammack_2009}.

The heating induced by the "writing" beam is clearly signaled by the
photoluminescence dynamics after the laser excitation. Precisely,
we show in Fig. S3.D the time resolved emission at the position of the "writing"
beam (red) and outside (blue): at the position of
the laser excitation the photoluminescence emission is enhanced by $\approx$ 45
$\%$ at the end of the "writing" pulse while the photoluminescence enhancement
is much weaker for the region outside of the ``writing`` beam. This non-linear
response marks
the collapse of the inner-ring at the end of the laser excitation. It signals
the
efficient thermalization of indirect excitons after the ''writing'' pulse. Thus,
the
population of optically active states is enhanced and therefore the
photoluminescence emission
is sharply increased. Finally, from these measurements we also deduce the
optical lifetime of
indirect excitons:  
the photoluminescence 
obeys a
mono-exponential decay after the laser excitation (black line in Fig. S3.D),
with
a time constant equal to
138(3) ns.

\vspace{0.5cm}
\centerline{\includegraphics[width=8.5cm]{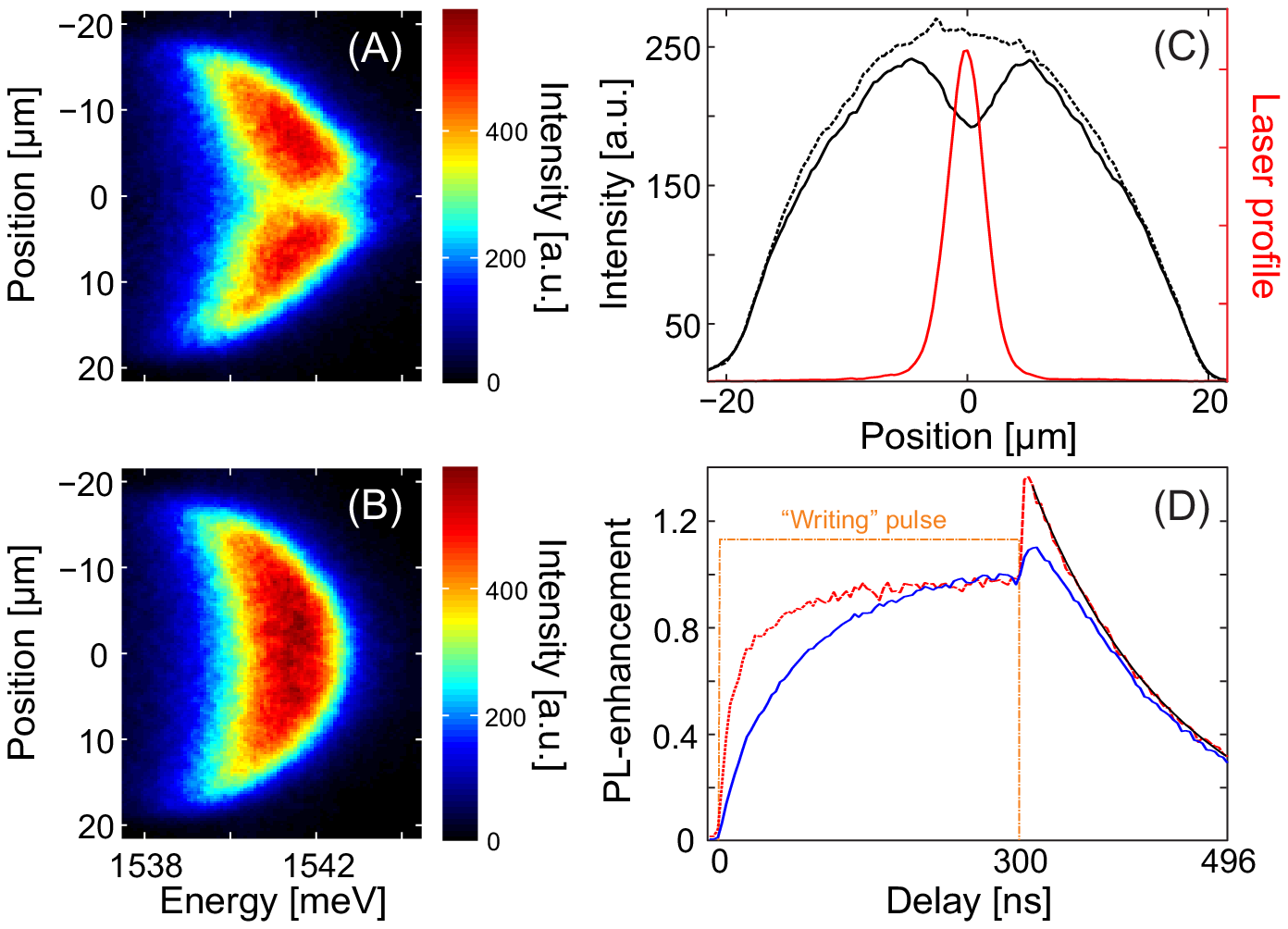}}
\noindent {\bf Fig. S3.} (A-B): Spatially and spectrally resolved emission of
indirect
excitons at the end (A) and 4 ns after (B) a ''writing'' pulse
 with 6 $\mu$m waist and at 10 W.cm$^{-2}$. (C): Emission profiles at the end of
the writing pulse and 4 ns later, solid and dotted lines respectively 
(the laser profile is shown by the red line). (D): Dynamics of the
photoluminescence at the position of the "writing" pulse (red) and outside
(blue). The black line
shows the best fit which yields a mono-exponential decay with a time constant
equal to 138(3) ns. All measurements were realized at 350 mK.

\section{Exciton concentration in the hollow-trap}

In Figure S4, we present the spatially and spectrally resolved emission of
indirect excitons confined in a hollow-trap for the same experimental
conditions as for Figure 3.C. We note that the
photoluminescence inside the trap is shifted to higher energies, by
$\delta E$, than outside the trap where the exciton gas
is more dilute. This difference results from the repulsive dipole-dipole
interactions
between indirect excitons and the energy
shift, $\delta E$, gives an estimation of the gas density inside the trap.
Indeed, the mean field energy associated to repulsive
exciton-exciton interactions may be expressed as
$u_0.n_\mathrm{X}$, where $u_0$ is
a constant factor controlled by the DQW geometry and the correlations
between excitons while $n_\mathrm{X}$ denotes the
exciton density \cite{Schindler_08,Rapaport_09}. For the experiments displayed
in Figure S4 we have $\delta E\approx$ 1.3 meV which corresponds to
$n_\mathrm{X}\approx$ 5 10$^{10}$ cm$^{-2}$
\cite{Schindler_08,Rapaport_09,Ben_Tabou_2001}.

\vspace{0.5cm}
\centerline{\includegraphics[width=5.5cm]{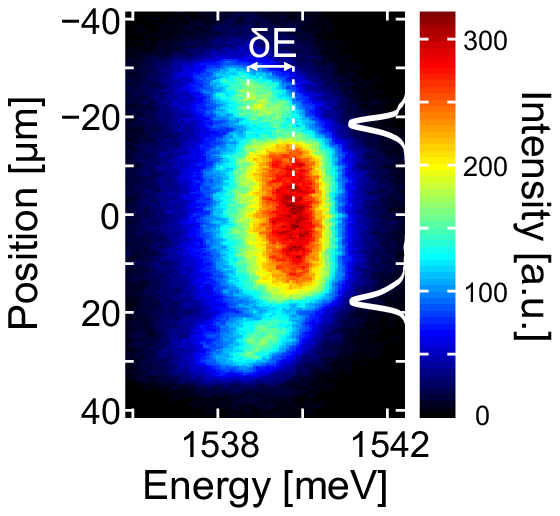}}
\noindent {\bf Fig. S4.} Spatially resolved photoluminescence spectrum along the
central
axis of the hollow-trap. These measurements were realized in the same
experiments as for Fig. 3.C. The white line displays the spatial profile of the
``writing`` beam which defines the trap barriers.
$\delta E\approx$ 1.3 meV is the energy shift of the emission from which the
exciton
concentration is deduced.

\section{Pulse sequences}

In our experiments, indirect excitons
were optically created by two CW diode lasers (DL-100 from Toptica) at
640nm (``preparation`` beam) and 790nm (``writing`` and probe beams). Both
lasers were
pulsed with electro-optic (integrated-optical modulator from JenOptic) and
acousto-optic
modulators (Crystal Technologies Inc.), controlled by a pulse sequencer. Thus,
we engineered up to four independent excitation
pulses with rising and falling edges of $\sim$2ns and $\sim$5ns for the
electro-optic and
acousto-optic switching respectively. The measurements that we report were
performed using 4
different pulse sequences which are resumed in Figure S5.

\vspace{0.5cm}
\centerline{\includegraphics[width=8.5cm]{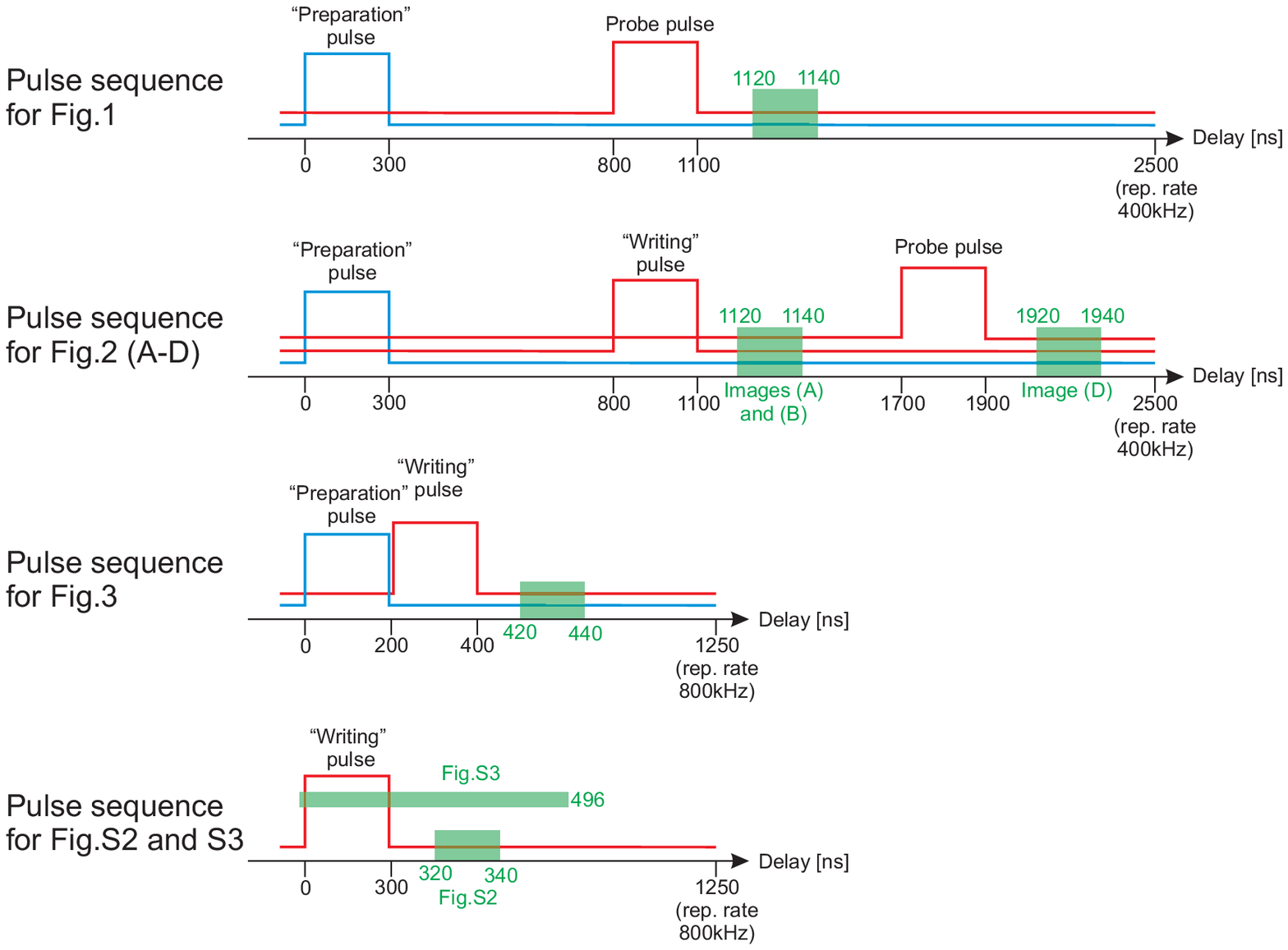}}
\noindent {\bf Fig. S5.} Blue and red-colored pulses correspond  to
``preparation`` and ``writing`` beams respectively. The
green-colored region provides the details of the time window during which real
images and spectrally
resolved ones were acquired.


\end{document}